\documentclass[fleqn,10pt]{wlscirep}
\usepackage[utf8]{inputenc}
\usepackage[T1]{fontenc}

\usepackage[whole]{bxcjkjatype}

\usepackage{graphicx}

\usepackage[nameinlink]{cleveref}
\usepackage{url}
\usepackage{color}
\usepackage{xcolor}
\usepackage{hyperref}
\hypersetup{
colorlinks = true, 
urlcolor = blue, 
linkcolor = blue, 
citecolor = blue 
}

\usepackage{multirow}
\usepackage{booktabs}

\definecolor{myblue}{RGB}{33,150,243}
\definecolor{mygreen}{RGB}{76, 175, 80}
\definecolor{purple}{RGB}{170, 0, 255}
\definecolor{satred}{RGB}{177,156,217}
\definecolor{satblue}{RGB}{119,158,203}

\usepackage{ulem}
\usepackage{bm}

\usepackage{soul}

\def\Wmat{\bm W}
\usepackage{lineno}
\usepackage{letltxmacro}
\LetLtxMacro{\originaleqref}{\eqref}
\renewcommand{\eqref}{Eq.~\originaleqref}

\newcommand{\journal}[1]{{\it #1}}
\title{Detecting anomalous citation groups in journal networks}

\author[a]{Sadamori Kojaku}
\author[b,c]{Giacomo Livan}
\author[d,e,f,*]{Naoki Masuda}

\affil[a]{Luddy School of Informatics, Computing, and Engineering. Indiana University, Bloomington, Indiana, 47408, USA}
\affil[b]{Department of Computer Science, University College London, London, WC1E 6EA, UK}
\affil[c]{Systemic Risk Centre, London School of Economics and Political Science, London, WC2A 2AE, UK}
\affil[d]{Department of Mathematics, University at Buffalo, State University of New York, Buffalo, New York, USA}
\affil[e]{Computational and Data-Enabled Science and Engineering Program, University at Buffalo, State University of New York, Buffalo, New York, 14260-2900, USA}
\affil[f]{Faculty of Science and Engineering, Waseda University, Tokyo, 169-8555, Japan}
\affil[*]{naokimas@buffalo.edu}

\begin{abstract}
The ever-increasing competitiveness in the academic publishing market incentivizes journal editors to pursue higher impact factors.
This translates into journals becoming more selective, and, ultimately, into higher publication standards.
However, the fixation on higher impact factors leads some journals to artificially boost impact factors through the coordinated effort of a ``citation cartel'' of journals.
``Citation cartel'' behavior has become increasingly common in recent years, with several instances being reported.
Here, we propose an algorithm----named CIDRE---to detect anomalous groups of journals that exchange citations at excessively high rates when compared against a null model that accounts for scientific communities and journal size.
CIDRE detects more than half of the journals suspended from Journal Citation Reports due to anomalous citation behavior in the year of suspension or in advance.
Furthermore, CIDRE detects many new anomalous groups, where the impact factors of the member journals are lifted substantially higher by the citations from other member journals.
We describe a number of such examples in detail and discuss the implications of our findings with regard to the current academic climate.

\end{abstract}
\begin{document}

\flushbottom
\maketitle
%
%
\thispagestyle{empty}

\section*{Introduction}

The volume of published research is growing at exponential rates \cite{Bornmann2015growth}, creating a pressing need to devise fast and fair methods to evaluate research outputs.
Measuring academic impact is a controversial and challenging task \cite{barnes2017h}.
Yet, the evaluation of research has increasingly been operationalized in terms of the citations received by research papers and citation-based bibliometric indicators, such as the $h$-index and the journal impact factor (JIF), which are widely used to evaluate individual researchers, academic institutions, and the research output of entire nations \cite{Garfield1992,Adam2002,King2004,Bornmann2015}.

Editors and academic publishers are under increasing pressure to ensure that their journals achieve and sustain high values of JIF \cite{garfield2006history} and other bibliometric indicators.
Such indicators are widely recognized as proxies for a journal's quality and prestige \cite{saha2003impact}, and have a considerable impact on the journal's readership numbers and subscription base.
This fact, in turn, incentivizes editors to devise strategies aimed at increasing citation numbers.
Such strategies may ultimately result in publications of a higher quality. However, there have been multiple reports of malicious practices merely aimed at boosting citation numbers.

Editors of some journals have generated citations for their journals by coercing the authors of submitted papers \cite{Wilhite2012} or by writing editorial reviews \cite{Foo2011}.
Such self-citations are relatively easy to spot because they involve only one journal.
Concerns have grown for less detectable forms of manipulation which involve the coordinated effort of a number of journals, a practice known as citation cartels.
Such a practice---also referred to as citation stacking---consists of groups of journals exchanging citations at excessively high rates \cite{Franck1999,Fister2016}.
For example, one instance of a citation cartel attracted attention in 2011.
In this example, two papers published in different journals provided a number of citations to a single journal, increasing its JIF by 25\%  \cite{Davis2011}.
Since then, new instances of citation stacking have been reported every year \cite{VanNoorden2013,Davis2014,TitleSuppressions}.

Journal editors may set up citation cartels by informally agreeing with other journal editors and colleagues to coerce citations \cite{VanNoorden2013,RetractionWatch2017}.
Such citation cartels are easy to launch and hard to detect.
To tackle this issue, Journal Citation Reports (JCR), which is owned by Clarivate Analytics and was owned by Thomson Reuters until 2016, has excluded from its annual journal ranking some \textit{pairs} of journals when at least one of the two journals cited the other excessively and the journal pair satisfied some additional criteria~\cite{JCR2019}.
As of 2019, JCR has suspended from its annual journal ranking 46 pairs of journals----featuring 55 journals in total---due to excessive pairwise citations \cite{TitleSuppressions,JCR2019}.
Alternatively, a previous study regarded citation cartels as groups of densely interconnected nodes (i.e., communities) in journal citation networks \cite{Fister2016}.
However, the approach based on network communities may suffer from false positives because communities are the norm rather than the exception in journal citation networks: journals tend to cite other journals in the same research field, which forms densely connected communities \cite{Lancichinetti2012,Hric2018,Rosvall2008}.

Detecting anomalous citation groups is inherently challenging because the abundance and prominence of citation communities may overshadow the anomalous citation patterns.
To address this challenge, we propose the CItation Donors and REcipients (CIDRE) algorithm.
The key idea of CIDRE is to discount the amount of citations between communities using a null model of networks with communities.
The null model accounts for the citation rates that can be expected under healthy citation practices due to journals' (\textit{i}) proximity (in terms of research areas) and (\textit{ii}) size (in terms of citation volumes, both given and received).
Then, CIDRE finds groups of journals with excessive within-group citations relative to the null model.
To the best of our knowledge, no empirically validated tool has been proposed for identifying \textit{groups} of journals whose citation practices can be regarded----with statistical confidence---as anomalous.

We apply the algorithm to a citation network of 48,821 journals across various disciplines constructed from Microsoft Academic Graph \cite{Sinha2016}.
CIDRE detects more than half of the instances suspended from JCR in the year of suspension or earlier.
Furthermore, CIDRE identifies a number of additional anomalous journal groups, including 7 groups in 2019 whose journals received more than 30\% of their incoming citations from other members of the group.
In the absence of a ground truth validation---such as the one provided by comparisons against the list of journals banned in JCR---we shall refrain from identifying these groups as citation cartel candidates, and it is clear that, in some cases, the anomalous citation patterns are not a result of citation cartels.
However, through extensive examples we will demonstrate that these groups are interpretable and composed of different patterns of anomalous citation behaviors.

Our results reveal a large number of journals that receive a disproportionate amount of their citations from a tiny group of publication venues, which account for a substantial fraction of these journals' impact factors (in excess of $50\%$ in some cases). In our final remarks, we will discuss how these findings should encourage a critical approach to the use of bibliometric indicators.
The Python code for CIDRE is available at GitHub~\cite{cidrecode}.

\section*{Results}

\subsection*{Data}

We use a snapshot of Microsoft Academic Graph (MAG) released on January 30th, 2020 to construct citation networks of journals \cite{Sinha2016}.
The data set contains bibliographic information including citations among 231,926,308 papers published from 48,821 journals in various research fields.
The bibliographic information includes the journal name, publication year, references, and author names.
We construct a directed weighted network of journals for each year $t$ between 2000 and 2019, in which a node represents a journal, and an edge indicates citations between journals.
We define the weight $W_{ij}$ of the edge from journal $i$ to journal $j$ in year $t$ by the number of citations from papers published in $i$ to papers published in $j$ made in the time window used for calculating the JIF, i.e., last two years $t' \in [t-2, t-1]$.
We use the term effective citation to refer to a citation reflected in the calculation of the JIF (i.e., a citation to a paper published in the last two years).
Unless stated otherwise, the citations in the following text refer to effective citations.

\subsection*{Detecting anomalous citation groups}

In citation networks, a citation cartel is manifested as a group of journals that excessively cite papers published in other journals within the group.
Although not all such groups are necessarily citation cartels, we aim to identify journal groups with excessive within-group citations.
Specifically, we assume that an anomalous citation group is composed of donor journals and recipient journals.
A donor journal provides excessive citations to papers published in recipient journals in the previous two years i.e., the time window for the JIF.
In cases where two journals exchange citations at excessively high rates, they simultaneously behave as both donors and recipients.
Although donor journals have no apparent direct benefit in providing citations to recipient journals, we consider them as a member of the anomalous citation group because some previously identified instances contain journals giving excessive citations to particular journals, which often share the publishers or editors \cite{Davis2011,VanNoorden2013,Davis2014}.

We identify excessive citations between journals using a null model for citation networks.
Specifically, we use the degree-corrected stochastic block model (dcSBM) \cite{Karrer2011,Peixoto2017} as the null model.
The dcSBM generates randomized networks that preserve the number of citations between groups of journals (i.e., blocks), and the outgoing and incoming citations of each journal on average.
We determine the blocks by fitting the dcSBM using a non-parametric Bayesian method \cite{Peixoto2017}.
Community detection methods for networks including the dcSBM have been shown to provide a reasonable partitioning of journal citation networks into research fields \cite{Lancichinetti2012,Hric2018,Rosvall2008}.
Therefore, the networks generated by the dcSBM are considered to be random networks that roughly preserve the patterns of citations within and across research fields.

CIDRE removes from the given network all the edges that are statistically compatible with the null model and then computes a donor score and a recipient score for all journals based on the residual edges in the network (see the Materials and Methods section).
In the following, we refer to the weights of such edges as excessive citations.
Consider a journal group, denoted by $U$, that contains journal $i$.
Journal $i$'s donor score, denoted by $x_{\text{d}}(i, U)$, is the fraction of excessive citations that journal $i$ provides to the other journals in $U$.
Journal $i$'s recipient score, denoted by $x_{\text{r}}(i, U)$, is the fraction of excessive citations that $i$ receives from other journals in $U$.
CIDRE considers a journal as a donor journal and a recipient journal if $x_{\text{d}}(i, U)$ and $x_{\text{r}}(i, U)$ are larger than a prescribed threshold $\theta = 0.15$, respectively (see the Discussion section for the choice of the $\theta$ value).

To find anomalous citation groups, CIDRE initializes $U$ to be the set of all nodes in the network.
Then, CIDRE removes from $U$ the journals that are neither a donor nor a recipient and recomputes the donor and recipient scores for the journals remaining in $U$.
CIDRE iterates the removal of nodes and recomputation of scores until no journal is further removed.
We partition $U$ into disjoint groups $U_{\ell}$ ($\ell=1,2,\ldots$), where each
$U_\ell$ is the maximal weakly connected component in the network consisting of nodes belonging to $U$ and the residual edges.
We regard each weakly connected component $U_{\ell}$ with more than $\theta_{\text{w}} = 50$ within-component citations as an anomalous citation group.

\subsection*{Overlap with the journal groups suspended from JCR}

\begin{figure}
    \centering
    \includegraphics[width=0.8\textwidth]{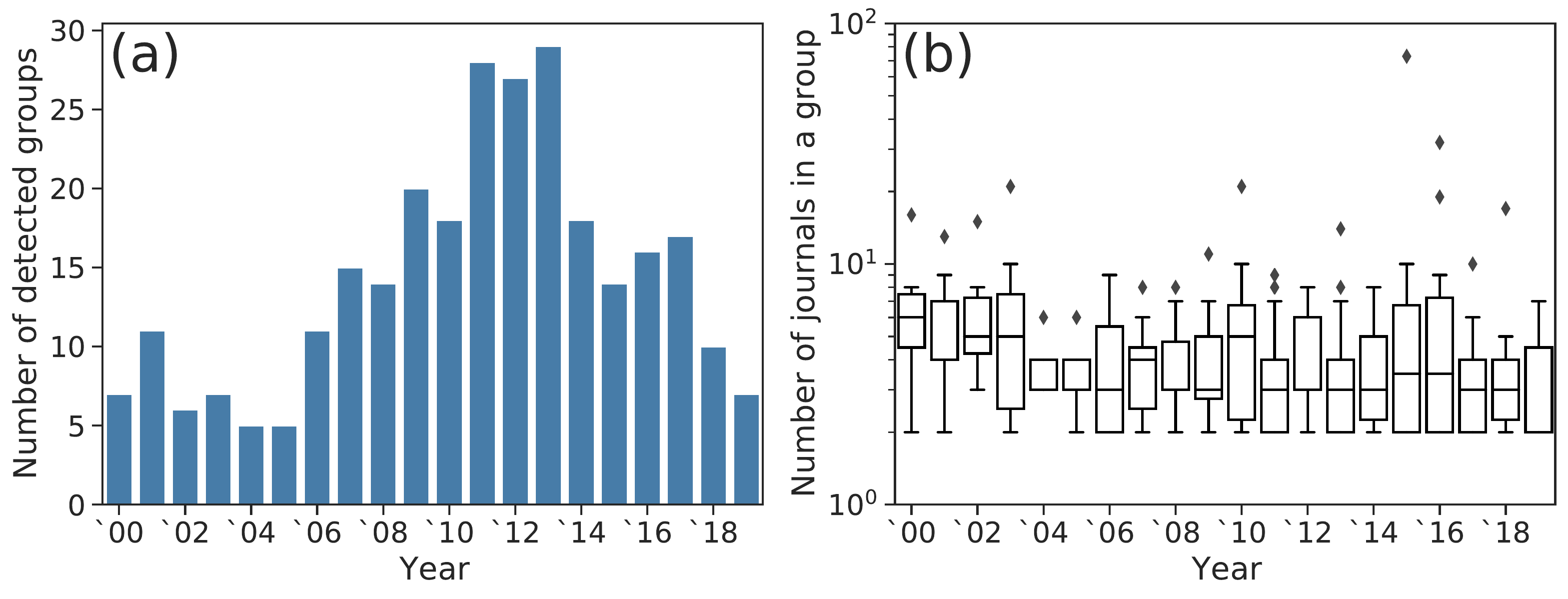}
    \caption{
    Statistics of anomalous citation groups detected by CIDRE.
    (a) Number of anomalous citation groups.
    (b) Number of journals in a group.
    A diamond indicates an outlier that does not fall in the range in $[q_{0.25}-1.5\Delta_{\text{IQR}}, q_{0.75}+ 1.5\Delta_{\text{IQR}}]$,
    where $q_{0.25}$ and $q_{0.75}$ are the first and third quartiles, respectively, and $\Delta_{\text{IQR}}=q_{0.75} - q_{0.25}$.
    }
    \label{fig:cartel-stat}
\end{figure}

CIDRE detected 184 citation groups between years 2010 and 2019 (Fig.~\ref{fig:cartel-stat}(a)). A detected citation group consisted of four journals on average (Fig.~\ref{fig:cartel-stat}(b)).
Because no ground truth is available for evaluating the detected groups, we compare them with the journals suspended from JCR.

Since 2007, JCR has suspended 227 journals due to excessive citations, of which 173 journals are suspended due to excessive self-citations, 55 journals due to excessive citations between two journals, and one journal due to both self-citations and pairwise citations \cite{MasterJournalList2019}.
Although JCR does not disclose its precise algorithm, they have released some criteria for suspensions.
Their criteria include the fraction of citations that the recipient journal receives from the donor journal, akin to the recipient score, together with the year since the first publications from the journals and the ranking of journals \cite{JCR2019}.
JCR reported 46 pairs of donor and recipient journals for excessive pairwise citations.
Some journal pairs suspended from JCR share a journal.
We merge such overlapping journal pairs suspended in year $t$ into one group, denoted by $U^{\text{JCR}} _{\ell}$, and
consider that $U^{\text{JCR}} _{\ell}$ is identified by JCR in year $t-1$ (i.e., one year prior to the suspension).
There are 22 such groups, which we denote by J1, J2, $\ldots$, J22.

\begin{figure}
    \centering
    \includegraphics[width=\textwidth]{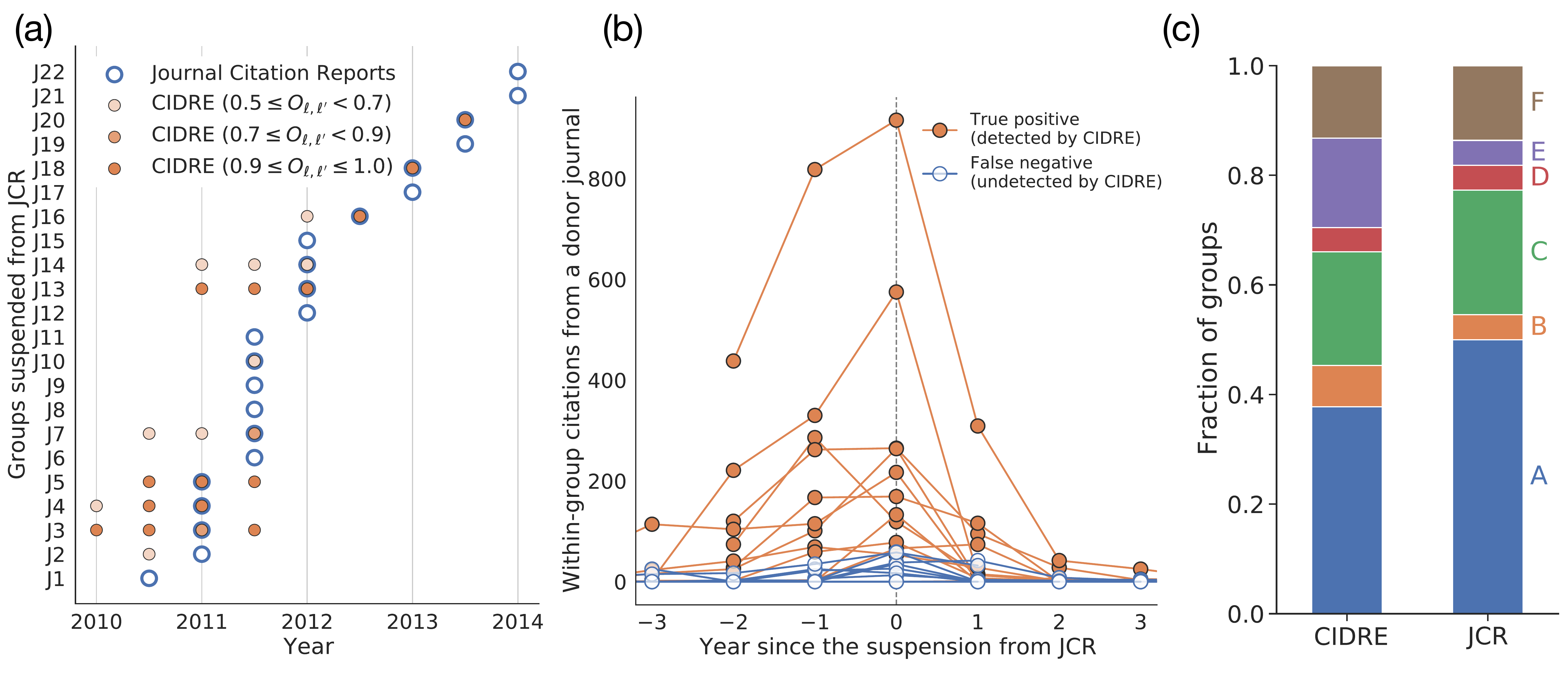}
    \caption{
        Overlap between journal groups identified by JCR and CIDRE.
        (a) Years when the journal groups are identified.
        The circles with thick borders represent the groups suspended from JCR.
        The filled circles represent those detected by CIDRE.
        The hue of the circles indicates the value of the overlap, $O_{\ell, \ell'}$, i.e., the fraction of journals in group $\ell'$ detected by CIDRE that belong to group $\ell$ suspended from JCR.
        CIDRE detected 11 out of the 22 groups suspended from JCR in the year of the suspension or before.
        (b) Number of within-group citations per year from a donor journal excluding self-citations.
        If the group has multiple donor journals, we show the count for the donor journal that provides the largest number of within-group citations.
        The horizontal axis indicates the year relative to that in which JCR suspended the journal group.
        Therefore, a negative value indicates a year before JCR suspended the journal group.
        For groups J17, J19, and J22, the donor journals do not provide any within-group citations for the three years before the suspension. (c) Classification of journal groups identified by CIDRE and JCR.
    }
    \label{fig:vs-thomson-reuters}
\end{figure}

We calculate the overlap between groups reported in JCR and CIDRE as $O= \big| U^{\text{JCR}}\bigcap U^{\text{CI}}\big| \big/ | U^{\text{CI}} \big|$, where $U^{\text{CI}}$ is a set of journals in a group detected by CIDRE.
If $U^{\text{JCR}}$ and $U^{\text{CI}}$ have $O \geq 0.5$ and share at least two journals,
we say that $U^{\text{JCR}}$ is detected by CIDRE.
CIDRE detects the 12 groups suspended from JCR at least once, of which 8 groups have $O \geq 0.8$ (Fig.~\ref{fig:vs-thomson-reuters}(a)).
CIDRE detects 10 groups earlier than JCR reports.
Furthermore, CIDRE detects 7 groups for multiple years before the suspension by JCR but no group after one year from the suspension, suggesting that they stopped malicious citation practices after the suspension had been lifted.

Could the above suspended groups also be detected by standard community detection algorithms?
To address this question, we consider three community detection algorithms, i.e., the modularity maximization by the Leiden algorithm~\cite{Traag2019}, Infomap~\cite{Rosvall2008}, and the dcSBM~\cite{Peixoto2017}.
We apply the algorithms and evaluate whether or not the detected communities match the suspended groups under the same matching criteria used for testing CIDRE.
These community detection algorithms have found at least three times more groups than those found by CIDRE.
However, none of them matches the suspended groups, with a small overlap of $O_{\ell, \ell'}\leq 0.012$ for all detected communities.
One may argue that the groups suspended from JCR---which consist of less than five journals---are too small to be detected with these community detection algorithms.
We have therefore run another experiment by restricting the number of nodes in each community (i.e., community size).
Specifically, the Leiden algorithm and the dcSBM accept a parameter with which one can control the community size.
We set the maximum community size to five for the Leiden algorithm and the average community size to three for the dcSBM.
We again find that no community matches the suspended groups, i.e., $O_{\ell, \ell'}\leq 0.087$ for all detected communities.
These results support that anomalous citation groups are difficult to detect with community detection algorithms.

Why are some suspended groups not detected by CIDRE?
The groups identified by JCR but not by CIDRE have considerably fewer within-group citations than the groups identified by both (Fig.~\ref{fig:vs-thomson-reuters}(b)).
Notable examples are groups J17, J19, and J22.
In these groups, the donor journals identified by JCR did not provide any within-group citations.
The lack of within-group citations is due to the fact that MAG is curated by a machine learning algorithm which sometimes fails to parse citations and publications, particularly for retracted papers~\cite{visser2019large,Martin-Martin2020}.
For instance, JCR suspended the journals comprising group J1 due to the anomalous citations from two papers published in the donor journal~\cite{Davis2011}.
Two papers were later retracted and not indexed in the MAG.
If we add back the retracted papers and rerun CIDRE, then CIDRE detects group J1.

\subsection*{Newly detected citation groups in 2010--2018}

\begin{figure}
    \centering
    \includegraphics[width=\hsize]{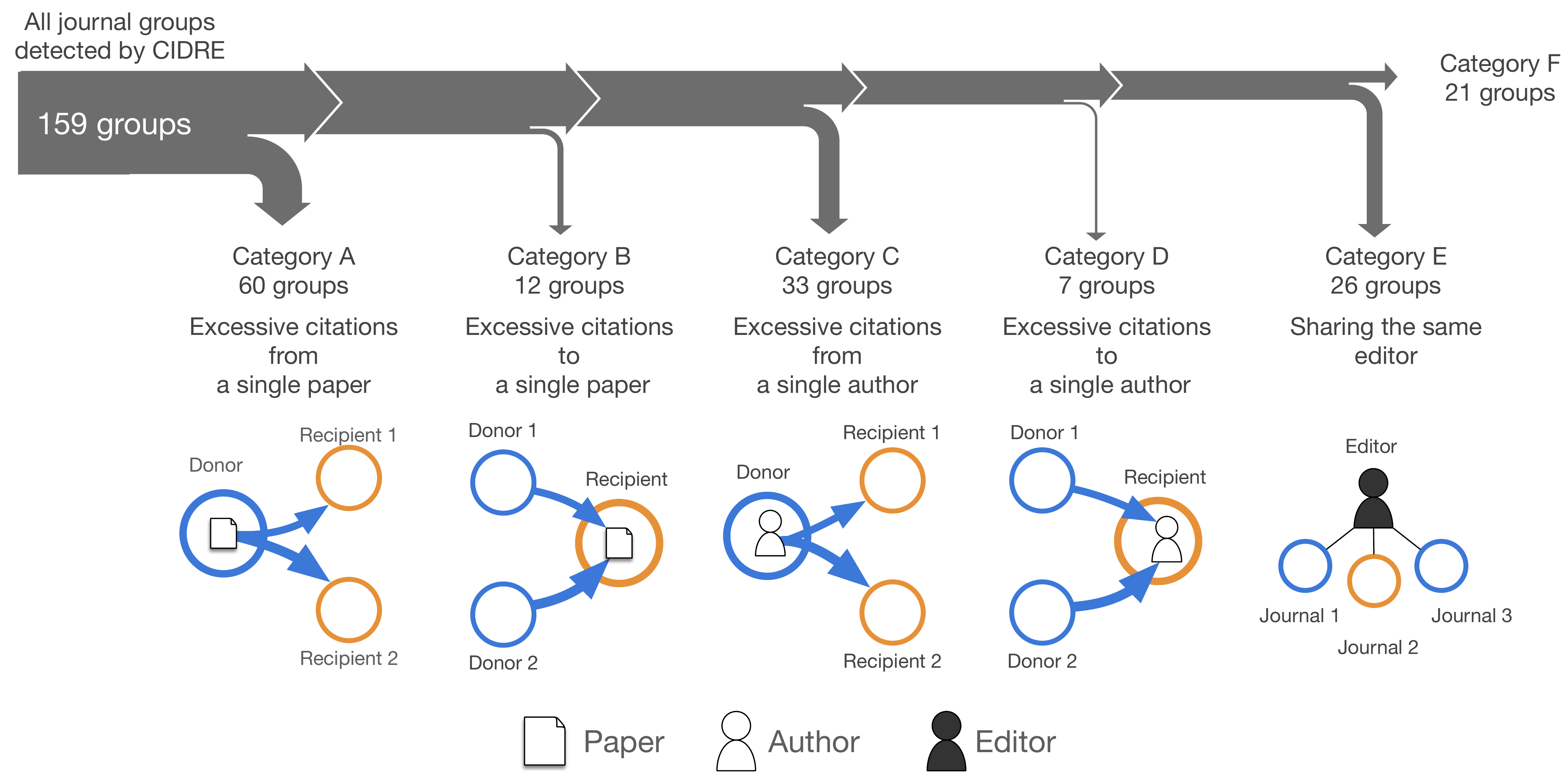}
    \caption{%
        Sankey diagram for the classification of groups. We sequentially apply five classification criteria based on citation patterns (A--D) or the overlap of editorial board members (E).
        Groups are classified into categories A--D if more than 20\% of within-group citations (A) come from a single paper, (B) go to a single paper, (C) come from a single author, or (D) go to a single author.
        For the groups that do not meet any of the criteria A--D, we classify a group into category E if its member journals have at least one overlapping editor. Otherwise, we classify it into category F.
    }
    \label{fig:schematic-classification}
\end{figure}

CIDRE detected 159 groups that JCR has not suspended.
We classified these detected groups based on five criteria that we sequentially applied:
more than 20\% of within-group citations (A) come from a single paper, (B) go to a single paper, (C) come from a single author, or (D) go to a single author, or (E) two journals in the group share at least one editorial board member (Fig.~\ref{fig:schematic-classification}; see Methods section for the method for identifying editors).
Over half of the groups detected by CIDRE (93 groups; 58\%) are attributed to excessive citations provided by a single paper (category A) or single author (category C).
In 19 groups (12\%), excessive citations are directed to a single paper (category B) or a single author (category D).
In 26 groups (16\%), journals share at least one editorial board member (category E).
The remaining 21 groups (13\%) do not meet any of the five criteria (category F).
For comparison, we apply the same classification rule to the 22 groups of journals suspended from JCR (Fig.~\ref{fig:vs-thomson-reuters}(c)).
Similar to the CIDRE, relatively many groups that are suspended from JCR belong to category A or C.
In the following, we closely inspect the groups with the largest number of within-group citations in each category except category E for which we inspect the group with a large overlap of editorial board members across the member journals.

\begin{figure}
    \centering
    \includegraphics[width=0.8\hsize]{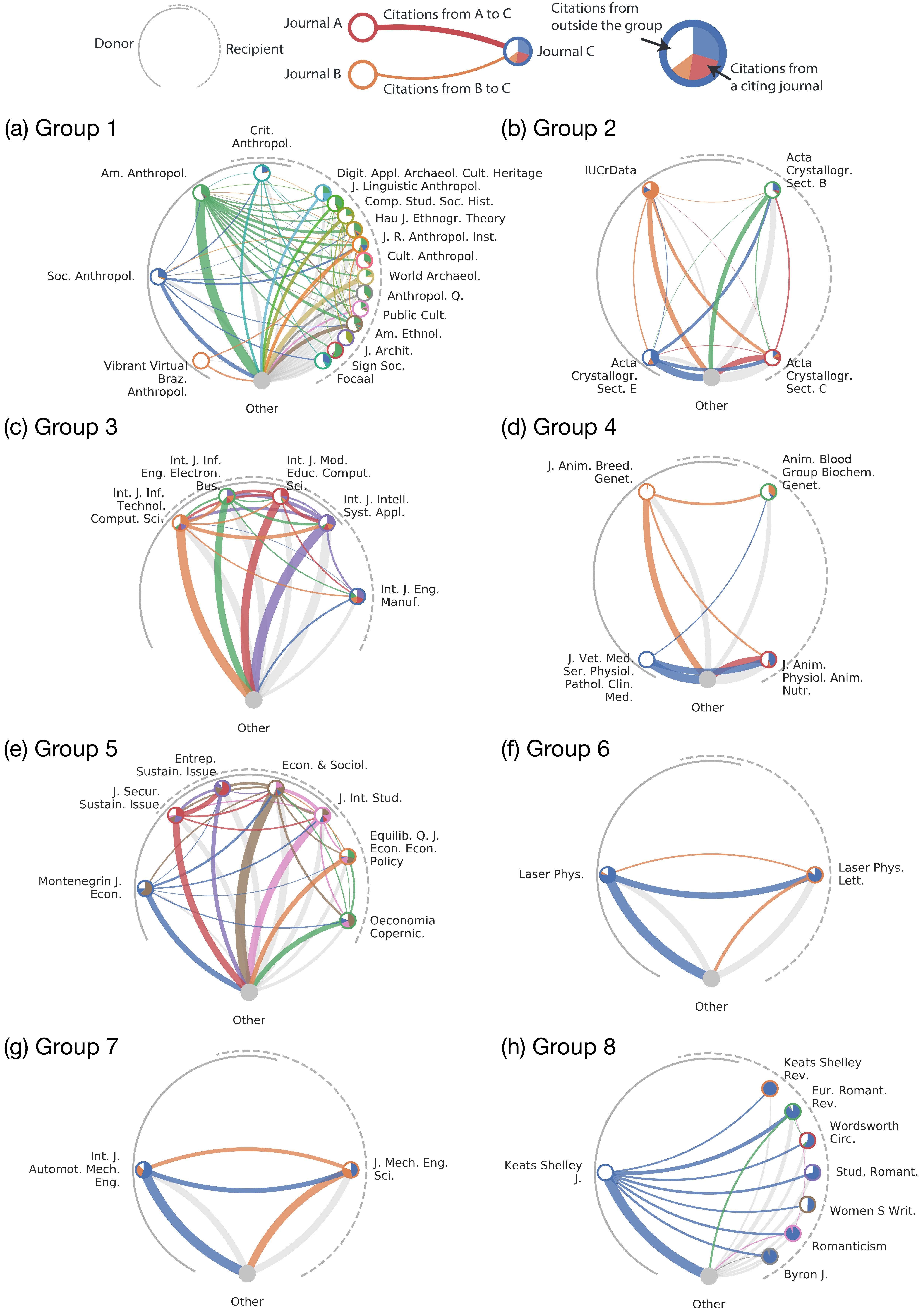}
    \caption{%
        Citation groups detected by CIDRE.
        The circles indicate journals.
        The color of edges indicates the source of citations.
        The width of edges is proportional to the number of citations made between two journals.
        Self-citation edges are omitted.
        The pie in color within each circle indicates the share of citations from the citing journal from the viewpoint of the cited journal.
        The white pie within each circle indicates the share of citations from outside the group.
        The journals on the left and right arcs indicate the donor and recipient, respectively.
        The journals that are simultaneously donor and recipient are located where the arcs overlap.
        The journals outside the group are agglomerated into a gray-colored circle at the bottom of each panel.
    }
    \label{fig:cartels}
\end{figure}

An instance of category A is group 1, which CIDRE detected in the network in 2018 and is composed of 17 journals on anthropology (Fig.~\ref{fig:cartels}(a)).
Two review papers published in donor journals, \journal{American Anthropologist}  and \journal{Social Anthropology}, provided 233 citations in total to the journals in group 1, of which 230 citations (99\%) were made to the papers published in the JIF time window.
Removing the citations from the two review papers decreases the JIFs for the 4 recipient journals, \journal{Anthropological Quarterly}, \journal{Cultural Anthropology}, \journal{Focaal}, and \journal{Journal of the Royal Anthropological Institute}, by more than 26\%.

An instance of category B is group 2, which CIDRE detected in the network in 2017 and is composed of four journals on crystallography (Fig.~\ref{fig:cartels}(b)).
Most of the within-group citations were made to a single paper published in a recipient journal, \journal{Acta Crystallographica Section C} (denoted by $R_{2,1}$).
In fact, the paper received 594 citations from the two donor journals, \journal{IUCrData} (denoted by $D_{2,1}$) and \journal{Acta Crystallographica Section E} (denoted by $D_{2,2}$), which accounts for 94\% of citations that $R_{2,1}$ received from $D_{2,1}$ and $D_{2,2}$.
Removing the within-group citations to the single paper decreases the JIF of $R_{2,1}$ by 22\%.
The paper is titled ``Crystal structure refinement with SHELXL'', which describes a software commonly used in crystallography.
The donor journals, $D_{2,1}$ and $D_{2,2}$, required the software users to cite the paper in their submission guidelines.

An instance of category C is group 3, which CIDRE detected in the network in 2014 and is composed of five journals on engineering (Fig.~\ref{fig:cartels}(c)).
Most of the within-group citations are attributed to self-citations across different journals by a single author (Fig.~\ref{fig:cartels}(c)).
The author wrote approximately one-third of papers (23 out of 63 papers) contributing to the within-group citations.
These papers provided 313 citations to the author's papers published in the recipient journals in 2012 and 2013.
The author was on the editorial board for \journal{International Journal of Intelligent Systems and Applications}, which serves as both a donor and recipient journal in this group.

An instance of category D is group 4, which CIDRE detected in the network in 2010 and is composed of four journals on veterinary science  (Fig.~\ref{fig:cartels}(d)).
One author wrote 33 papers published in a donor journal, \journal{Journal of Veterinary Medicine} (denoted by $D_{4,1}$), in 2010.
These papers provided 74 citations to 45 papers written by the same author published in a recipient journal, \journal{Journal of Animal Physiology and Animal Nutrition} (denoted by $R_{4,1}$), in 2008 and 2009.
The self-citations made by the author accounts for 35\% of citations that $R_{4,1}$ receives from $D_{4,1}$.

An instance of category E is group 5, which CIDRE detected in the network in 2016 and is composed of seven journals on business (Fig.~\ref{fig:cartels}(e)).
There are 176 editorial board members in total that serve any of the member journals. Among them, 19 individuals were the editors of at least two member journals.
Nearly half of the overlapping editors (9 out of 19) serve two journals, \journal{Journal of Security and Sustainability Issues} (JSSI) and \journal{Entrepreneurship and Sustainability Issues} (ESI), which account for at least 25\% of the editorial board members in each of the two journals.
The editor-in-chief of ESI, who also serves JSSI as an editor, provided and received the largest number of citations (62 and 55 citations, respectively) among the authors of papers published in this journal group.

An instance of category F is group 6, which CIDRE detected in the network in 2011 and is composed of two journals on laser science (Fig.~\ref{fig:cartels}(f)).
The donor journal, \journal{Laser Physics}, provided 1984 citations to the recipient journal, \journal{Laser Physics Letters}.
We did not find any concentration of citations; neither a single paper nor a single author provided or received more than 8\% of citations within the group.
In 2011, the number of citations from the donor journal to the recipient journal increased more than double, from 987 citations in 2010 to 1984 citations in 2011.
CIDRE identified the increase in the citations to be excessive and detected this group.

In addition to groups 1--6, two citation groups caught our attention, which we refer to as groups 7 and 8.
Group 7 is present in the network in 2017.
This group belongs to category E and consists of two journals on engineering (Fig.~\ref{fig:cartels}(g)).
CIDRE detected this group in five successive years between 2013 and 2017.
In 2017, a donor journal, \journal{International Journal of Automotive and Mechanical Engineering}, provided 81 citations to a recipient journal, \journal{Journal of Mechanical Engineering and Sciences}, of which 53 (65\%) citations were provided to the papers published in the JIF time window. Removing the 53 citations decreases the JIF of the recipient journal by 46\%.
The donor and recipient journals have 32 and 30 editorial board members, respectively, of which three editors overlap.
The two of the overlapping editors serve as the advisor and associate editor of both journals.

Group 8 is present in the network in 2016.
This group belongs to category A and consists of eight journals on literature (Fig.~\ref{fig:cartels}(h)).
A donor journal, \journal{Keats Shelley Journal}, provided 119 citations to the seven recipient journals, of which 110 (92\%) citations were provided to the papers published in the JIF time window.
Removing the 119 citations decreases the JIF of the recipient journals by at least 57\%.
A single paper titled ``Annual Bibliography for 2015'' provided all the within-group citations from the donor journal to the recipient journals.
This paper consists of 42 pages, of which 40 pages are the reference list.
In each of year 2012, 2013, and 2014, the donor journal published a paper with a similar title (e.g., ``Annual Bibliography for 2014'') that cited many papers in the recipient journals.
In these years, CIDRE detected the groups that consisted mostly of the donor journal and the recipient journals in group 8.

\subsection*{Anomalous journal groups in 2019}

\begin{figure*}
    \centering
    \includegraphics[width=0.8\hsize]{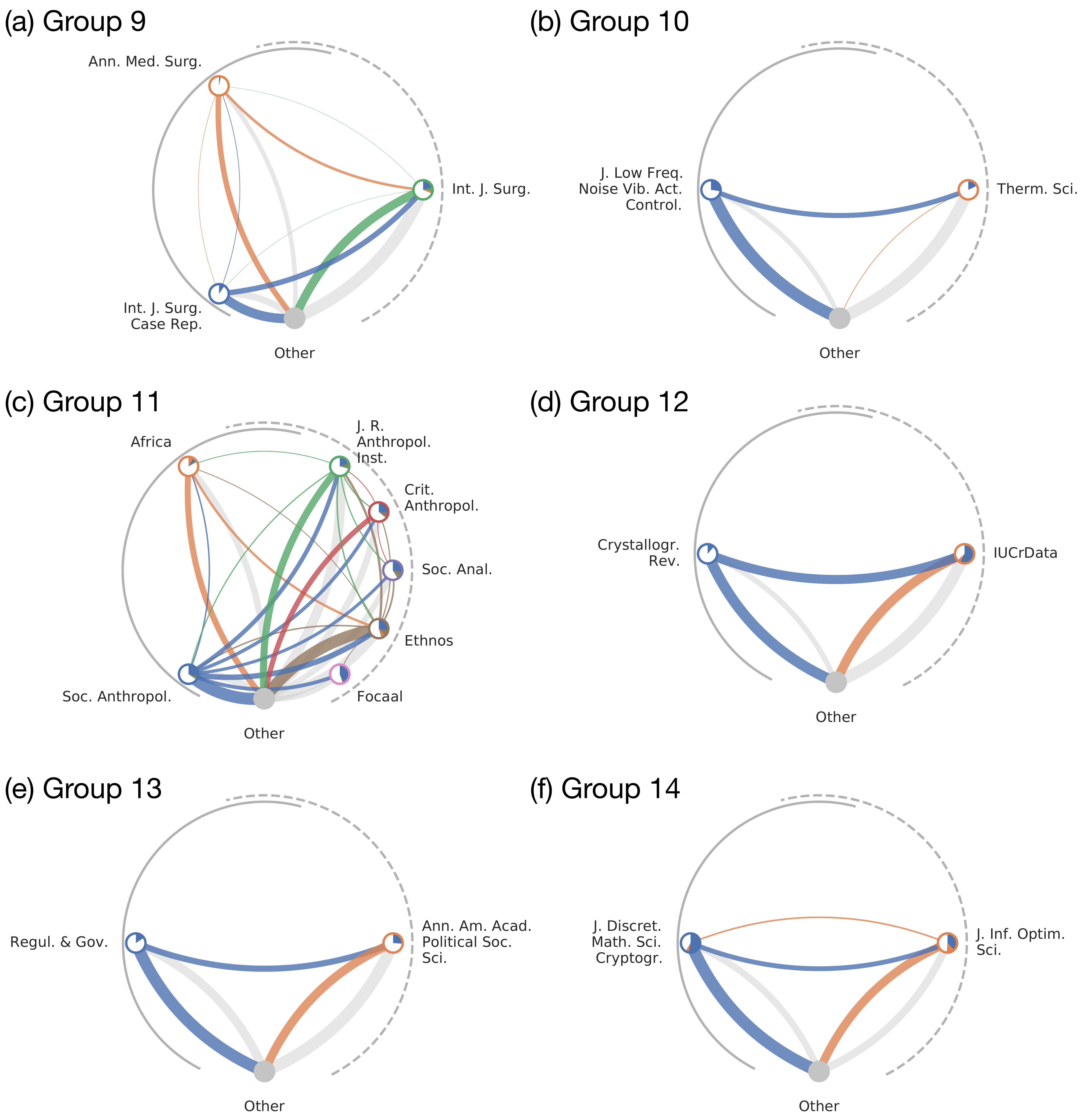}
    \caption{%
        Citation groups that CIDRE detected for the network in 2019.
    }
    \label{fig:cartels-SI}
\end{figure*}

While JCR has not reported any anomalous pairwise citations in the 2019 citation data, CIDRE detected 7 citation groups in 2019.
Six out of these seven groups belong to category A, B, C, D, or E (Fig.~\ref{fig:cartels-SI}).
We refer to these six journal groups as groups $9, 10, \ldots, 14$.

Group 9 consists of three journals on surgery and belongs to category B (Fig.~\ref{fig:cartels-SI}(a)).
As is the case for group 2, most of the within-group citations pointed to a single paper published in the sole recipient journal, \journal{International Journal of Surgery} (denoted by $R_{9}$).
The paper received 483 citations from the two donor journals, \journal{Annals of Medicine and Surgery} (denoted by $D_{9,1}$) and \journal{International Journal of Surgery Case Reports} (denoted by $D_{9,2}$), which account for 82\% of the citations (i.e., 592) that $R_{9}$ receives from $D_{9,1}$ and $D_{9,2}$.
Removing these citations decreases the JIF of $R_{9}$ by 20\%.
The paper is titled ``The SCARE 2018 statement: Updating consensus Surgical CAse REport (SCARE) guidelines,'' which is a guideline for surgical reports.
In the guideline for the authors, the donor journals request the authors to cite the paper as a condition for submission.
Furthermore, the author of the SCARE paper is the managing and executive editor of $D_{9,2}$ and $R_{9}$.
In addition to this editor, $D_{9,2}$ and $R_{9}$ share many editors.
In fact, there are 107 and 84 editors in $D_{9,2}$ and $R_{9}$, respectively, of which 79 individuals are the editors of both journals.
Journals $D_{9,1}$, $D_{9,2}$, and $R_9$ conducted a similar citation practice in the previous two years.
In fact, CIDRE detected a group composed of $D_{9,2}$ and $R_{9}$ in 2018, in addition to the present group in 2019.
In 2017 and 2018, $D_{9,1}$ and $D_{9,2}$ requested the authors to cite the previous version of the SCARE guideline paper written by the same author published in $R_9$  in 2016.
There were 559 and 554 citations from the donor journals to the paper in 2017 and 2018, respectively.
The new guideline paper entered the time window for the JIF when the old guideline paper exited the time window.

Group 10 is composed of two journals and belongs to category D (Fig.~\ref{fig:cartels-SI}(b)). The donor journal, ``\journal{Journal of Low Frequency Noise, Vibration and Active Control}'', provided 160 citations to the recipient journal, \journal{Thermal Science}.
A single author received 74 out of the 160 citations (46\%) from 34 papers published in the donor journal.
The 29 out of the 34 papers are included in a special issue of which the author was the guest editor.
The special issue consists of 74 papers.

Group 11 is composed of seven journals on anthropology and belongs to category A (Fig.~\ref{fig:cartels-SI}(c)).
A single review paper published in a donor journal, \journal{Social Anthropology}, cited 95 papers published in the recipient journals, all of which were published in the time window for the JIF.
If one removes the citations from that review paper, the JIF of each of the five recipient journals decreases by more than 18\%.
In the review paper, the author acknowledged the editors of the two recipient journals, \journal{Social Analysis} and \journal{Focaal}, owned by a publisher, Berghahn Journals, for granting access.

Group 12 consists of two journals on crystallography and belongs to category A (Fig.~\ref{fig:cartels-SI}(d)).
A single paper published in the donor journal, \journal{Crystallography Reviews}, cited 124 papers published in the recipient journal, \journal{IUCrData}, all of which were published in the time window for the JIF.
If one removes these 124 citations, the JIF of the recipient journal decreases by 57\%.

Group 13 is composed of two journals on political science and belongs to category D (Fig.~\ref{fig:cartels-SI}(e)).
The donor journal, \journal{Regulation and Governance}, provided 95 citations to the recipient journal, \journal{Annals of American Academy of Political and Social Science}.
Removing the citations from the donor decreases the JIF of the recipient by 26\%.
The 89 out of 95 (93\%) citations from the donor to recipient journals pointed to the papers included in a special issue of the recipient journal, i.e., ``Regulatory Intermediaries in the Age of Governance.''
The special issue consists of 16 papers, each of which received less than 2 citations on average from journals outside group 11 in 2019.
The special issue was edited by three guest editors who are on the editorial board of the donor journal.
The three editors wrote a paper in the special issue.
The paper received 26 citations in 2019, of which 13 citations (50\%) came from the donor journal.
The paper was highlighted as the most cited paper in the last three years in the recipient journal in 2019.

Group 14 consists of two mathematical journals and belongs to category E (Fig.~\ref{fig:cartels-SI}(f)).
The donor journal, \journal{Journal of Mathematical Sciences and Cryptography}, published 150 papers, of which 52 papers cited 36 papers published in the recipient journal, \journal{Journal of Information and Optimization Science} in the time window for the JIF.
We did not find a single author or a single paper that was exclusively cited or was cited within the group.
The 52 papers published in the donor journal were written by 126 authors, of which 107 authors (84\%) had never cited the recipient journal before.
Both donor and recipient journals have the same chief editor.

\section*{Discussion}

In this paper, we put forward an algorithm---named CIDRE---to identify groups of journals that cite each other at excessively high rates.
CIDRE detects a majority of journal groups suspended from JCR.
Notably, in several cases, it does so years in advance.
In addition, it detects a number of anomalous groups, whose members increased their JIFs by 17--130\% via within-group citations.
The inspection of such groups reveals a variety of mechanisms leading to such inflation.
Specifically, more than half of the anomalous groups are due to one paper or one author that singlehandedly provides or receives many citations within the group.

The algorithm's practical value lies in that it is deterministic and scalable to large networks, which makes it possible to apply it in an online fashion to incoming streams of new citation data.
Furthermore, it can be applied to different types of networks.
For instance, CIDRE could be applied to bipartite author--journal networks, where a directed edge indicates a publication by an author in the journal, in order to detect potential predatory practices, such as the publication of papers with little peer review \cite{beall2017learned}.
CIDRE could also be applied in different contexts, e.g., to detect the manipulation of ratings in e-commerce platforms and social media \cite{livan2017excess}.

One should be careful when drawing conclusions from the application of CIDRE.
The comparison against the ground-truth data provided by JCR, and the manual inspection of the groups detected by CIDRE support that the groups flagged by CIDRE warrant consideration as potential citation cartel candidates.
That being said, we ought to acknowledge that some of such candidates may arise due to unintended biases such as geographical proximity \cite{Katz1994,Pan2012}, reciprocity between peers \cite{Li2019}, and editorial preferences \cite{Yoon2013,Perez2019}, rather than to outright malicious citation practices.
In this respect, CIDRE should not be considered as a tool for automated decision-making or a substitute for expert judgment, but rather a support tool to extract interpretable information from the complexity of journal citation networks.

CIDRE has a parameter---the threshold $\theta$---that sets the minimum fraction of excessive citations that the donor/recipient journals provide/receive within their group.
Changing the value of $\theta$ induces a hierarchical onion-like structure on the detected journal groups.
The inner cores that survive with a larger $\theta$ value are considered to be tighter citation groups, which may be more plausible citation cartel candidates.
In this study, we set $\theta = 0.15$ to allow for a fair comparison with JCR; all recipient journals suspended from JCR received at least 15\% of their incoming citations from donor journals \cite{Davis2017}.
Then, we manually inspected each group detected by CIDRE to pinpoint individual papers, authors, editors, and specific journals associated with excessive citations.
However, manual inspection is a costly task and hard to scale up when dealing with large numbers of groups.
This problem will manifest itself when one analyzes citation groups composed of authors because an author network can be much larger than a journal network.
Therefore, in practice, it may be useful to prioritize journal groups that survive with higher thresholds.
With CIDRE, one can easily determine the ranking of journal groups according to this criterion because gradually increasing $\theta$ to reveal onion-like structure is straightforward and not computationally too costly.

Regardless of the conclusions that one may draw on specific anomalies, our findings reveal the widespread presence of journals whose JIFs are substantially hoisted by the citations received from a small group of other journals.
It would be hard not to relate this with the ever-increasing emphasis on citations and bibliometric indicators, and the pressure it puts on journal editors to boost growth in such numbers.
We believe our findings to be a rather direct consequence of this environment, where actors are incentivized to act on the very same metrics according to which they are ranked, in a feedback loop that closely echoes Goodhart's Law: ``when a measure becomes a target, it ceases to be a good measure'' \cite{manheim2018categorizing}.
In this respect, we believe that our results should encourage a more critical and nuanced approach to the use and interpretation of citation-based bibliometric indicators.

\section{Methods}

\subsection*{Detection of anomalous citation groups}
We assume that an anomalous citation group is composed of journals that act as donors, recipients, or both.
A donor journal gives excessive citations to the journals in the same group.
A recipient journal receives excessive citations from the journals in the same group.

Algorithm CIDRE finds groups of journals, $U$, composed of the donor and recipient journals.
We quantify the extent to which a journal $i$ acts as donor or recipient within group $U$ using the donor score $x_{\text{d}}$ and the recipient score $x_{\text{r}}$, respectively.
They are defined by
\begin{linenomath*}
    \begin{align}
        x_{\text{d}}(i, U) & := \frac{1}{s^{\text{out}} _ i} \sum_{j \in U, j\neq i} W_{ij} h(i,j), \\
        x_{\text{r}}(i, U) & := \frac{1}{s^{\text{in}} _ i} \sum_{j \in U, j\neq i} W_{ji} h(j,i),
    \end{align}
\end{linenomath*}
where $s^{\text{out}}_i:=\sum_{j=1} ^N W_{ij}$ and $s^{\text{in}}_i:=\sum_{j=1} ^N W_{ji}$ are the out-strength and in-strength of journal $i$, respectively, and
$N$ is the number of nodes.
Function $h(i,j)$ is an indicator function, where we set $h(i,j) = 1$ if citations from journal $i$ to journal $j$ are excessive relative to the null model; otherwise $h(i,j)=0$.
The donor and recipient scores range in $[0,1]$.
A large donor score for journal $i$, i.e., $x_{\text{d}}(i, U)$, implies that $i$ cites papers in other journals in $U$ more often than expected under a null model; similar for the recipient score.

The citations from journal $i$ to journal $j$ are deemed to be excessive if and only if they satisfy the following two conditions.
First, more than half of citations made to papers published in any previous years from $i$ to $j$ were made to papers published in the last two years (i.e., effective citations).
Second, the number of citations, $W_{ij}$, is larger than that expected for a null model.
Specifically, for each directed edge from node $i$ to node $j$, we compute the $p$-value as the probability $p_{ij}$ that the null model assigns a weight $w$ that is larger than or equal to the actual weight of edge $(i,j)$ in the given network, i.e., $W_{ij}$. One obtains
\begin{linenomath*}
    \begin{align}
        p_{ij} = 1 - \sum_{w=0} ^{W_{ij}-1} P^{\text{null}}_{ij}\left( w; \hat \lambda_{ij} \right), \label{eq:pvalue}
    \end{align}
\end{linenomath*}
where $\hat \lambda_{ij}$ is a parameter for the null model.
We describe the null model in the next section.

We perform a statistical test for each edge at the significance level of $\alpha=0.01$, with the Benjamini-Hochberg correction \cite{Benjamini1995} to suppress the false positives due to the multiple comparison problem.
In other words, one regards $m$ edges with the smallest $p$-values as significant (i.e, $h(i,j)=1$) and other edges as insignificant (i.e., $h(i,j)=0$).
The number $m$ is given by the largest integer $\ell$ for which $p^{(\ell)} \leq \ell \alpha/M$, where $p^{(\ell)}$ is the $\ell$th smallest $p$-value and $M$ is the number of edges in the network.

After removing the insignificant edges, we seek groups of journals that have a donor or recipient score larger than a prescribed threshold $\theta$.
To this end, we use the following algorithm, akin to the $k$-core decomposition algorithm \cite{Batagelj2010}.
First, we prune the network by keeping only the edges with $h(i,j)=1$.
Second, we initialize $U=\{1,\ldots, N\}$, and compute the donor and recipient scores for each node.
Third, we remove a node $i$ from $U$ if $x_{\text{d}}(i,U)< \theta$ and $x_{\text{r}} (i,U) < \theta$.
Then, we recompute the donor and recipient score for all neighbors of $i$.
We repeat the third step until no node is removed.
Fourth, we partition $U$ into disjoint groups $U_{\ell}$ ($\ell=1,2,\ldots$), where each $U_{\ell}$ is a maximal weakly connected component in the edge-pruned network composed of the nodes in $U$.
We expect that anomalous citation groups contain sufficiently many within-group citations.
Therefore, we remove $U_{\ell}$ if the sum of the weight of edges within $U_{\ell}$ except self-loops is less than $\theta_{\text{w}}$.
We set $\theta = 0.15$ and $\theta_{\text{w}} = 50$.
We note that CIDRE is a special case of the generalized core decomposition algorithm \cite{Batagelj2010} with vertex property function $f(i,U)=\max(x_{\text{d}}(i,U), x_{\text{r}}(i,U))$.

\subsection*{Null model}

We employ the dcSBM \cite{Karrer2011,Peixoto2017} as a null model.
The dcSBM consists of blocks, where each block is a group of journals.
The dcSBM places an edge from node $i$ to $j$ ($i,j=1,2,\ldots, N$) with a probability determined by the block memberships, out-strength $s^{\text{out}}_i$ of node $i$ in the original network, and in-strength $s^{\text{in}} _j$ of node $j$.
The generated networks preserve the expectation of $s^{\text{out}}_i$ and $s^{\text{in}} _i$ for each node $i$, and the expected number of edges between and within the blocks of the given network.

With the dcSBM, one assumes that the weight of the edge from node $i$ to $j$ obeys a Poisson distribution given by \cite{Karrer2011}
\begin{linenomath*}
    \begin{align}
        P_{ij} ^{\text{null}} (w; \lambda_{ij}) = \frac{\lambda_{ij} ^w \exp(-\lambda_{ij})}{w!},
    \end{align}
\end{linenomath*}
where $P_{ij} ^{\text{null}}(w;\lambda_{ij})$ is the probability that the dcSBM assigns weight $w$ ($w = 0,1,2,\ldots$).
Parameter $\lambda_{ij}$ is equal to the mean for the Poisson distribution, i.e., the expected number of citations for the null model.
We set $\lambda_{ij}$ to the maximum likelihood estimator conditioned on the blocks, which is given by
\begin{linenomath*}
    \begin{align}
        \label{eq:lambda_mle}
        \lambda_{ij} = \frac{s^{\text{out}}_{i} s^{\text{in}}_{j} \Lambda_{g_i, g_j}}{S^{\text{out}}_{g_i}S^{\text{in}}_{g_j}},
    \end{align}
\end{linenomath*}
where $g_i$ is the ID of the block to which node $i$ belongs, $\Lambda_{uv}$ is the number of directed edges from block $u$ to block $v$,
$S^{\text{out}}_u = \sum_{\ell=1}^N s^{\text{out}}_{\ell} \delta(g_{\ell}, u)$ and $S^{\text{in}}_u = \sum_{\ell=1}^N s^{\text{in}}_{\ell} \delta(g_{\ell}, u)$ are the sum of out-strength and in-strength of the nodes in block $u$, respectively, and $\delta(\cdot, \cdot)$ is Kronecker delta \cite{Karrer2011}.

One may be tempted to use the $\lambda_{ij}$ value given by \eqref{eq:lambda_mle} to compute the $p$-value using \eqref{eq:pvalue}.
However, if $\lambda_{ij}$ is smaller than one, even the edges with the smallest weight $W_{ij}=1$ may be judged to be excessive in the significance test explained in the previous section.
We instead require $W_{ij}$ to be large for journal $i$ to be regarded to excessively cite journal $j$.
Therefore, we use a clipped value, $\hat \lambda_{ij}$, to compute the $p$-value using \eqref{eq:pvalue}, where
\begin{linenomath*}
    \begin{align}
        \hat \lambda_{ij} = \max(1, \lambda_{ij}).
    \end{align}
\end{linenomath*}

We find the blocks by fitting the dcSBM to the journal citation networks.
Specifically, we first construct an aggregated network, in which the weight of the edge from node $i$ to node $j$, denoted by $\overline W_{ij}$, is given by
the sum of the weight over the networks between 2000 and 2019, i.e., $\overline W_{ij} = \sum_{t = 2000}^{2019} W^{(t)}_{ij}$, where
$W^{(t)}_{ij}$ is the weight of the edge from node $i$ to node $j$ in the network in year $t$.
Then, we identify the blocks of the aggregate network using a non-parametric Bayesian method without hierarchical structure \cite{Peixoto2017}.
Note that we use the aggregated network $\overline \Wmat$ to find the blocks of journals.
Then, with the detected blocks, we compute $\lambda_{ij}$ given by \eqref{eq:lambda_mle} for each yearly network $\Wmat^{(t)}$.
This is because the number of citations monotonically increases over time.
Therefore, recent yearly citation networks tend to have more excessive citations than older networks if one uses $\lambda_{ij}$ computed for the aggregated network.

\subsection*{Identifying editorial board members}

There are 641 journals in the 184 groups detected by CIDRE.
We manually identified the web pages listing the editorial board members for 525 among the 641 journals.
Extracting human names, particularly non-Latin names, from web pages is challenging.
In addition, spelling variation makes it difficult to match editors in different journals.
Therefore, we did not aim to calculate the precise number of editorial board members shared by different journals but to calculate its lower bound.
Specifically, we extracted person names with the Spacy package~\cite{spacy}.
Then, one of the authors, S.K., manually inspected the extracted names, and removed non-human names and too short names (e.g., initials).
Using exact string matching for the manually inspected names, we matched the editors in different journals.

\section*{Author contributions}
N. M. conceived the research. S.~K. performed the numerical analysis. S.~K., G.~L., and N.~M. contributed to the development of the algorithm and wrote the manuscript.

\section*{Competing interests}
The authors declare no conflict of interest.

\section*{Data availability}

The data that are needed for reproducing the results are openly available in Microsoft Academic Graph at \url{https://academic.microsoft.com/home}.

\section*{Acknowledgements}
GL acknowledges support from an EPSRC Early Career Fellowship (Grant No. EP/N006062/1).
NM acknowledges support from AFOSR European Office (Grant No. FA9550-19-1-7024).

\bibliography{main}

\end{document}